\documentclass{article}
\usepackage[english]{babel}

\usepackage[letterpaper,top=2cm,bottom=2cm,left=3cm,right=3cm,marginparwidth=1.75cm]{geometry}

\usepackage{amsmath}
\usepackage{graphicx}
\usepackage[colorlinks=true, allcolors=blue]{hyperref}

\usepackage{amssymb}
\usepackage{hyperref}
\usepackage{listings}
\usepackage{mathtools}
\usepackage[utf8]{inputenc}
\usepackage[svgnames]{xcolor}
\usepackage{amsmath}
\usepackage{breqn}
\usepackage{textcomp}

\title{\textbf{Bogomolny approach in description \\
of superconducting structures}}
\author{{\L}ukasz T. St\c{e}pie\'{n}$^{1}$ \\ (1: \textit{The Pedagogical University of Cracow}) \\ \\ \& \\
\\
Krzysztof Pomorski$^{2,3}$ \\ (2: \textit{Cracow University of Technology}, 3: \textit{Quantum Hardware Systems, Lodz} )}

\begin{document}
\maketitle

\begin{abstract}
Bogomolny approach in the context of strong necessary conditions is formulated for the case of various superconducting and semiconductor structures.  Certain results were obtained for a constant magnetic field, which experimentally corresponds to spiral-like superconducting cable placed inside solenoid.
\end{abstract}

\section{Introduction to Bogomolny methodology}
Euler-Lagrange equations of many models in physics are nonlinear partial differential equations of second order, but sometimes
one can consider instead of them the equations of first order, the so-called Bogomolny equations (or Bogomol'nyi
  equations).
They were derived by Bogomolny for among others,
Ginzburg-Landau model in \cite{Bogomolny} (although independently, they were derived in \cite{BPST}, for another model - SU(2) Yang-Mills theory and then they are often called as BPS equations), similar results were obtained in \cite{Hosoya} (cited in this context, only in \cite{Bialynicki}). Some presentations of among others, the Bogomolny equations and their solutions for the Ginzburg-Landau model, were given for e.g. in  \cite{MantonSutcliffe} and \cite{Meissner2013}. The Bogomolny equations for
some modified Ginzburg-Landau model were derived in \cite{Contatto}, but there is a coupling between the Yang–Mills term in the Lagrangian with the Higgs field through a continuous function, so this has been a different model from this one investigated in this paper.
The models, for which one can derive Bogomolny equations (or BPS
equations), are called as BPS models. There are two advantages
of them. One can expect soliton solutions. Moreover, obtaining exact solutions is possible.
Such solutions significantly allow  for larger understanding of considered non-linear models. Bogomolny equations also guarantee existence of a topological Bogomolny bound, and this causes a topological stability of solitons, which carry a non-trivial value of the corresponding topological charge. Therefore, the BPS property
is very important.

Besides, there is a possibility of a reduction of the full second order static equation of motion to a set of first order equations.
The classical Bogomolny approach (completion to square in the action functional), was applied
in many papers, and was giving many important results.
However, a completely general method, which allows to derive
BPS equations (if possible), in a more systematic way. This method is called as the concept of strong necessary conditions (CSNC). It was originally introduced and analyzed in \cite{SOK1979}, \cite{SOK1981}, \cite{SOK1993}, \cite{SOK1a}, \cite{SOK2},
 \cite{SokStSok}, \cite{SOK2005}, \cite{SOK22}, \cite{SOK2a}.
\cite{SOK2009}, and it has been very recently further developed in 2016 by Adam and Santamaria, in \cite{Adam_Santamaria} (they proposed there the so-called first order Euler-Lagrange (FOEL) formalism). This concept was successfully applied for among others,
deriving Bogomolny equations for many nonlinear field-theoretical models, among others, the models from Skyrme-family models and gauged nonlinear $\sigma O(3)$ model, \cite{ST1}, \cite{ST2015},
\cite{ST11}, \cite{ST12}, \cite{ST3}.
In this paper, we apply the CSNC method to derive the Bogomolny equations (BPS equations), for the gauged Ginzburg-Landau model
in curved space.
This model was introduced by Ginzburg and Landau in the 1950's and this provides some description of superconductivity. Formally,
the G-L model is a limit of the BCS model (which explains
the phenomenon of superconductivity by the well-known notion
of Cooper pairs of superconducting electrons.
The G-L model has been investigated since many years, among others in
\cite{AkkermansMallick}, \cite{Albert}, \cite{BystrovEtAl}, \cite{Contatto},  \cite{ContiOttoSerfaty}, \cite{Jaykka},
\cite{PeninWeller}, \cite{Serfaty}, \cite{SandierSerfaty}, \cite{Weinan}, \cite{Yang}.

\section{The Bogomolny classical approach}
Let's present shortly the Bogomolny approach, basing on the
scalar field theory - model $\phi^{4}$ with spontaneous symmetry breaking

  \begin{equation}
  \begin{gathered}
  E=\int^{\infty}_{-\infty} \bigg( \frac{1}{2} \bigg(\frac{d \phi}{d x}\bigg)^{2} + \frac{\lambda}{2}(\phi^{2} - \gamma^{2})^{2} \bigg) dx,
  \label{fido4}
  \end{gathered}
  \end{equation}

  where $\phi(x) \in \mathbb{R}$, and $\lim_{x \rightarrow \pm \infty} \phi(x) = \pm \gamma$ (what can be seen as superconducting order parameter far from defect of superconducting order parameter that can take place in Josephson junction).
  The Euler-Lagrange equations for this model are similar to Ginzburg-Landau equation and have the form

  \begin{equation}
  \frac{d^{2} \phi}{d x^{2}} = 2\lambda \phi (\phi^{2} - \gamma^{2}).
  \end{equation}

   We may avoid solving of them, namely we write the formula for $E$ in (\ref{fido4}), as follows

  \begin{equation}
  \begin{gathered}
   E=\int^{\infty}_{-\infty} \bigg( \frac{1}{2} \bigg(\frac{d \phi}{d x} + \sqrt{\lambda}(\phi^{2} - \gamma^{2})\bigg)^{2} - \hspace{-0.05 in}
  \sqrt{\lambda} \hspace{0.08 in} \frac{d \phi}{d x} (\phi^{2} - \gamma^{2})  \bigg) dx,
  \label{fido4INN}
  \end{gathered}
  \end{equation}

   We integrate the term $\sqrt{\lambda} \hspace{0.08 in} \frac{d \phi}{d x} (\phi^{2} - \gamma^{2})$ in (\ref{fido4INN}). We get

  \begin{equation}
  \begin{gathered}
   E=\int^{\infty}_{-\infty} \frac{1}{2} \bigg(\frac{d \phi}{d x} + \sqrt{\lambda}(\phi^{2} - \gamma^{2})\bigg)^{2} dx +
   \frac{2\sqrt{\lambda}}{3} \gamma^{2} \mid Q \mid,\\
   Q = \phi(\infty) - \phi(-\infty),
  \label{fido4INNpowyc}
  \end{gathered}
  \end{equation}

  where $Q$ - topological charge . Topological charge can indicate presence of vortices in Josephson junction or just imprint phase on
  Josephson junction due to external magnetic field source or current flow via junction (defect of order parameter).
  If we now require reaching the minimum by the functional (\ref{fido4INNpowyc}), then the first term needs to vanish

 \begin{equation}
  \frac{d \phi}{d x} = \sqrt{\lambda}(\gamma^{2}-\phi^{2}). \label{rBgm}
  \end{equation}

  These is just the Bogomolny equation.

  The very-known solution of 
   is the so called "kink"

  \begin{equation}
  \phi(x) = \gamma \tanh{(\gamma \sqrt{\lambda} (x - x_{0}))}.
  \end{equation}
Kink solution can correspond to non-superconductor interface or system of two superconductors with different asymptotic phase imprint on macroscropic wave-function.
  \section{A short description of the Bogomolny equations for usual gauged Ginzburg-Landau model}

The Lagrangian for the standard gauged Ginzburg-Landau model
 is well-known and has the form, \cite{Meissner2013}

 \begin{equation}
          \begin{gathered}
	\mathcal{L}=- \frac{1}{4} F_{\mu \nu} F^{\mu \nu} - \bigg (\frac{\partial}{\partial x^{\mu}} + 2iA_{\mu} \bigg) \psi^{\ast}
	\bigg (\frac{\partial}{\partial x_{\mu}} - 2iA^{\mu} \bigg ) \psi - \frac{\beta}{4}(\mid \psi \mid^{2} - \eta^{2})^{2},
          \end{gathered}
	\end{equation}

   where $\mu = 0, 1, 2$ and $\eta^{2}=\frac{1}{8e^{2}\lambda^{2}}$ and $F_{\mu \nu}=\frac{\partial A_{\mu}}{\partial x^{\nu}} - \frac{\partial A_{\nu}}{\partial x^{\mu}}$ is anti-symmetric electromagnetic tensor, and $F_{{\mu \nu}} F^{\mu \nu}=\frac{1}{2}(B^2-\frac{E^2}{c^2})$. We consider static field configuration, then
          we have only magnetic field in $z$ direction and we focus on the Hamiltonian
          \begin{equation}
          \begin{gathered}
	\mathcal{H}= \mid \bigg(\bigg(\frac{\partial}{\partial x} - 2iA_{x}\bigg) \pm i \bigg (\frac{\partial}{\partial y} - 2iA_{y}
          \bigg)\bigg) \psi \mid^{2} +
          \frac{1}{2} (B_{z} \pm 2e(\mid \psi \mid^{2} - \eta^{2}))^{2} \pm \\
          2e\eta^{2} B_{z} + \frac{1}{4}(\beta - 8e^{2}) (\mid \psi \mid^{2} - \eta^{2})^{2},
            \end{gathered}
	\end{equation}

where $B_{z} = \frac{\partial A_{y}}{\partial x}- \frac{\partial A_{x}}{\partial y}$.

       This Hamiltonian can be written in the following form

             \begin{equation}
          \begin{gathered}
	\mathcal{H}= \mid \bigg(\bigg(\frac{\partial}{\partial x} - 2iA_{x}\bigg) \pm i \bigg (\frac{\partial}{\partial y} - 2iA_{y}
          \bigg)\bigg) \psi \mid^{2} +
          \frac{1}{2} (B_{z} \pm 2e(\mid \psi \mid^{2} - \eta^{2}))^{2} \pm 
          2e\eta^{2} B_{z}.
            \end{gathered}
	\end{equation}

 The most interesting case is, when: $\beta = 8e^{2}$. Then
 \begin{equation}
 H \geq 2e \eta^{2} \mid \int B_{z} d^{2}x \mid.
 \end{equation}

This inequality is saturated i.e. when ground state energy of macroscopic wave function attains minimum, when
 canonical momentum is set to zero, so we have
            \begin{equation}
          \begin{gathered}
             \bigg(\bigg(\frac{\partial}{\partial x} - 2iA_{x}\bigg) \pm i \bigg (\frac{\partial}{\partial y} - 2iA_{y}
          \bigg)\bigg) \psi = 0,  \label{rB1}\\
             B_{z} = \mp 2e(\mid \psi \mid^{2} - \eta^{2})
            \end{gathered}
	\end{equation}
         The last equation is value of critical magnetic field above which superconducting phenomena is vanishing. 	
          We call the equations 
           as Bogomolny equations (Bogomol'nyi equations, BPS equations).

         So, as we see in both examples: for the scalar field model and Ginzburg-Landau model,
          Bogomolny equations are the partial differential equations of first-order, in contrary to the Euler-Lagrange equations,
					which are (to this model), the partial differential equations of second-order.
					Besides, all solutions of Bogomolny equations, are also the solutions of the Euler-Lagrange equations (but the inverse
					situation does not hold in general).\\

          One can show that the flux of magnetic field as Abrikosov vortex is subjected to quantisation always:

          \begin{equation}
          \Phi = \int B_{z} d^{2} x = \frac{n \pi}{e},
          \end{equation}

          then
           \begin{equation}
					H \geq 2\pi \mid n \mid \eta^{2}
           \end{equation}

    (for the solutions of
     (\ref{rB1})
     we have the saturation of this inequality). So, we have a connection between minimal value of the energy of Coopair pair condensate and topological aspect of the solution.

      One applies the following ans\"{a}tzes for vector potential in 2 dimensions present in Abrikosov vortex with rotational symmetry given as
					
					\begin{gather}
					\psi(\vec{r}) = \eta e^{in\theta} (1+f(r)),\\
					A_{x} = -\frac{n}{e} \frac{y}{r^{2}}(1 + \alpha(r)),\\
                    A_{y} = \frac{n}{e} \frac{x}{r^{2}}(1 + \alpha(r)),\\
		       \end{gather}			

    where  $r^{2} = x^{2} + y^{2}$.				
					
					Then the magnetic field: $B_{z}=\frac{n \alpha'(r)}{2er}$ and the Bogomolny equations 
 are  \cite{Meissner2013}				
					\begin{gather}
					f'=-\frac{n\alpha(1+f)}{r}, \ \	\alpha'=-\frac{r}{2n\lambda^{2}} (2f + f^{2}). \label{rownania_na_alfa_f}
					\end{gather}
					
					We require certain asymptotic limits as superconducting order parameter in vortex core set to zero or constant flat superconducting order parameter far away from single vortex core that give us conditions:
     \begin{gather}
     \lim_{r \rightarrow 0} \alpha(r) = -1, \lim_{r \rightarrow 0} f(r) = -1,
					\lim_{r \rightarrow \infty} \alpha(r) = 0, \lim_{r \rightarrow \infty} f(r) = 0.
     \end{gather}

				No exact solutions of (\ref{rownania_na_alfa_f}) are known, however one can apply numerical procedures
        to find non-trivial and physically interesting solutions
         of (\ref{rownania_na_alfa_f}).
         For big $r$, we have asymptotically, \cite{Meissner2013}:

         \begin{gather}
         f(r) = -nK_{0}\bigg(\frac{r}{\lambda}\bigg) + O(e^{-2r/\lambda}), \\
				 \alpha = -\frac{r}{\lambda} K_{1}\bigg(\frac{r}{\lambda}\bigg) + O(e^{-2r/\lambda}),
         \end{gather}

         where $K_{0}, K_{1}$ are the modified Bessel functions of second kind. In this case:

         \begin{equation}
         2e\lambda^{2} B_{z} \rightarrow -nK_{0}\bigg(\frac{r}{\lambda}\bigg) + O(e^{-2r/\lambda}).
         \end{equation}

         For small $r$, we have \cite{Meissner2013}:

         \begin{gather}
         f(r) \approx -1 + C(n) r^{n}, \\
				 \alpha(r) \approx -1 + \frac{r^{2}}{4n\lambda^{2}},
         \end{gather}

         where $C(n)=const$ can be determined numerically for given $n$ that is integer number of fluxons inside Abrikosov vortex core,
         and $B_{z}$ tends to $B_{z}(0)=\frac{1}{4e\lambda^{2}}$.

\section{The gauged Ginzburg-Landau model in curved space and the concept of strong necessary conditions in equations of motion }

\begin{figure}
    \centering
    \includegraphics[width=8cm]{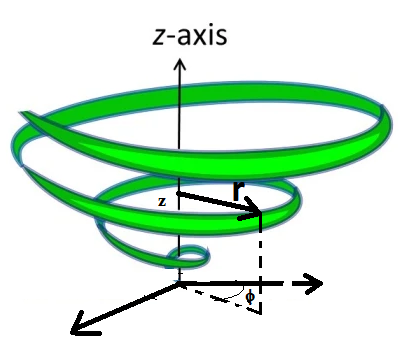}
    \caption{Case of curved superconducting cable parameterized in cylindrical coordinates by $\phi$ and z, so $r(z,\phi)$ as given by work on robust field-induced Josephson junction \cite{Pomorski}. }
    \label{fig:enter-label}
\end{figure}

Being motivated by work \cite{Curved} and \cite{Pomorski}  we formulate Langranian in curvilinear case referring to Fig.\ref{fig:enter-label}.

The Lagrangian of the explored gauged nonlinear Ginzburg-Landau model has the following form:

\begin{equation}
\begin{gathered}
\mathcal{L} = \frac{a_{1}(\phi, z) (-ih \omega_{,\phi} - \frac{eA_{\phi} \omega}{c}) (ih \omega^{\ast}_{,\phi} - \frac{eA_{\phi} \omega^{\ast}}{c})}{2m} + \frac{a_{2}(\phi, z) (-ih \omega_{,z} - \frac{eA_{z} \omega}{c}) (ih \omega^{\ast}_{,z} - \frac{eA_{z} \omega^{\ast}}{c})}{2m}+ \label{lagrGL} \\
\frac{a_{3}(\phi, z) (-ih \omega_{,\phi} - \frac{eA_{\phi} \omega}{c}) (ih \omega^{\ast}_{,z} - \frac{eA_{z} \omega^{\ast}}{c})}{2m} + \frac{a_{4}(\phi, z) (-ih \omega_{,z} - \frac{eA_{z} \omega}{c}) (ih \omega^{\ast}_{,\phi} - \frac{eA_{\phi} \omega^{\ast}}{c})}{2m} +\\
\frac{a_{5}(\omega, \omega^{\ast}) (-ih \omega_{,\phi} - \frac{eA_{\phi} \omega}{c})}{2m} + \frac{a_{6}(\omega, \omega^{\ast}) (-ih \omega_{,z} - \frac{eA_{z} \omega}{c})}{2m} +
\frac{a_{7}(\omega, \omega^{\ast}) (ih \omega^{\ast}_{,\phi} - \frac{eA_{\phi} \omega^{\ast}}{c})}{2m} +\\
\frac{a_{8}(\omega, \omega^{\ast}) (ih \omega^{\ast}_{,z} - \frac{eA_{z} \omega^{\ast}}{c})}{2m} + V(\omega, \omega^{\ast}) + c_{1}(A_{\phi,z} - A_{z,\phi})^{2},
\end{gathered}
\end{equation}

where $a_{i}=a_{i}(\phi, z), i=1, 2, 3, 4, a_{k} = a_{k}(\omega, \omega^{\ast}), k=5, 6, 7, 8$, $c_{1}=const$  and $\omega = \omega(\phi, z) \in \mathbb{C}$.
Here $V(\omega, \omega^{\ast}) $ encodes all superconducting properties incorporated in geometry of superconducting bended wire.  The coefficients $a_5, .. , a_8$ bring non-Hermicity to Hamiltonian and thus the described system can be categorized as dissipative. `
\section{The concept of strong necessary conditions in equations of motion}

The essence of the concept of strong necessary conditions is such that we replace considering of the Euler-Lagrange:
 equations,

 \begin{equation}
 F_{,u} - \frac{d}{dx}F_{,u_{,x}} - \frac{d}{dt}F_{,u_{,t}}=0, \label{el}
 \end{equation}

 following from the extremum principle, applied to the functional:

 \begin{equation}
 \Phi[u]=\int_{E^{2}} F(u,u_{,x},u_{,t}) \hspace{0.05 in} dxdt, \label{functional}
 \end{equation}

 with considering the strong necessary conditions, \cite{SOK1979}-\cite{SOK2009}

 \begin{gather}
   F_{,u}=0, \label{silne1} \\
   F_{,u_{,t}}=0, \label{silne2} \\
   F_{,u_{,x}}=0, \label{silne3}
 \end{gather}

 where $F_{,u} \equiv \frac{\partial F}{\partial u}$, etc.

 One can see easily that all solutions of the system of the equations (\ref{silne1}) - (\ref{silne3}), are the solutions of the Euler-Lagrange equation (\ref{el}). However, these solutions, if they exist, are very often trivial. Hence, we make gauge transformation of the functional (\ref{functional}):

  \begin{equation}
  \Phi \rightarrow \Phi + Inv, \label{gauge_transf}
  \end{equation}

 where $Inv$ is such functional that its local variation with respect to $u(x,t)$ vanishes:
 $\delta Inv \equiv 0$.\\
 Gauge transformation from \ref{gauge_transf} can correspond to change of system ground energy and one of example is placement of superconducting system inside solenoid with small magnetic field.

  Thanks to this feature, the Euler-Lagrange equations (\ref{el}) and the Euler-Lagrange equations resulting from  requiring of the extremum of $\Phi + Inv$, have the same form.
 On the other hand, the strong necessary conditions (\ref{silne1}) - (\ref{silne3}), are non-invariant with respect to the gauge transformation (\ref{gauge_transf}). Then, we have a chance to obtain non-trivial solutions. Obviously, the strong necessary
 conditions (\ref{silne1}) - (\ref{silne3}) constitute the system of the partial differential equations of the order less than the order of Euler-Lagrange equations (\ref{el}). \\
 Since we use here the notion of topological charge or topological invariant, the this is useful to say more about this issue.

 Let's take into account the sine-Gordon model (called also as  "sinus-Gordon model", and used for the description of the solitons in Josepshon conjunction), \cite{Felsager}:

 \begin{gather}
     \mathcal{L}  = -\frac{1}{2} \partial_{\alpha} u \partial^{\alpha} u - \frac{\mu^{2}}{\lambda^{2}} (1-\cos{(\lambda u)}).
 \end{gather}

Its symmetry is translational: $u \rightarrow u + \frac{2\pi}{\lambda} k, k \in \mathbb{Z}$ and one can label different sector in this model by a pair $(m,n)$, where $m, n \in \mathbb{Z}$
such that a field configuration, which belongs to $E_{n,m}$,
satisfies the following boundary conditions \cite{Felsager}
\begin{eqnarray}
\begin{gathered}
\lim_{x \rightarrow -\infty} u(x,t) = \frac{2\pi}{\lambda} n,\\
\lim_{x \rightarrow \infty} u(x,t) = \frac{2\pi}{\lambda} m.
\end{gathered}
\end{eqnarray}
Then, the asymptotic values of $u(x,t)$: $\lim_{x \rightarrow \pm \infty} u(x,t)$, are conserved, in other words they do not depend on time. We denote their difference
(for e.g. having meaning of phase drop across superconductor non-superconductor interface) by $Q_{topol}$, \cite{Felsager}:

\begin{equation}
Q_{topol} = u_{\infty} - u_{-\infty},
\end{equation}

and we interpret this as a conserved charge, so its density is: $\rho[u] = \frac{\partial u}{\partial x}$, and \cite{Felsager}

\begin{equation}
Q_{topol} = \int^{\infty}_{-\infty} \frac{\partial u}{\partial x} dx \equiv
\int^{\infty}_{-\infty} u_{,x} dx = u_{\infty} - u_{-\infty}
\end{equation}

The characteristic property of such charges is such that they are conserved independently on the dynamics.
This is the simplest situation, when the so-called homotopy group
is $\pi_{1}(S^{1})$. If the homotopy group is $\pi_{2}(S^{2})$, then the topological charge is (cf. \cite{Morandi}):

\begin{equation}
 Q_{topol} = \int^{\infty}_{-\infty} \int^{\infty}_{-\infty} (\omega_{,x} \omega^{\ast}_{,y}-\omega_{,y} \omega^{\ast}_{,x})dx dy,
\end{equation}

which after a generalization has the form (which was done in
\cite{SOK2}, \cite{SOK2a}):

\begin{equation}
 Q_{topol} = \int^{\infty}_{-\infty} \int^{\infty}_{-\infty} G_{1}(\omega,\omega^{\ast})(\omega_{,x} \omega^{\ast}_{,y}-\omega_{,y} \omega^{\ast}_{,x})dx dy.
\end{equation}

Now we explain, why such generalization is always important, when we apply strong necessary conditions. As we have written, the aim of this paper is to derive the Bogomolny decomposition (the Bogomolny equations), using strong necessary conditions, for Giznburg-Landau model with $U(1)$ gauge.
In the order to derive the Bogomolny decomposition, this is necessary to make the dual equations (following from strong necessary conditions) self-consistent. The generalizations of the topological charges (topological invariants) allow to choice properly the functions like $G(\omega, \omega^{\ast})$, in order to make the dual equations self-consistent. Thus, we have an opportunity to:

\begin{itemize}
\item make certain part of the dual equations linearly dependent – the remaining equations are just the Bogomolny equations
\item obtain the condition for the potential of the given field-theoretical model. The Bogomolny decomposition (the Bogomolny equations) exist only for this model, which potential satisfies such condition.
\end{itemize}

 An important issue is the issue of construction of the general form of the density of the topological invariant for the case of the topology of this model. Such invariant should be also a gauge-invariant term (respect to $U(1)$ gauge).
 A general construction of this density  was given (in the cases of the gauged models: nonlinear $\sigma \ O(3)$, restricted BPS baby Skyrme and full BPS baby Skyrme), in \cite{ST2015} (proposed there for the first time). At least one of gauge invariance density has the following form, \cite{ST2015}:

  \begin{equation}
 \begin{gathered}
 I_{1} =\lambda_{3} \cdot \bigg\{R'_{1} \cdot [i \cdot (\omega_{,\phi}\omega^{\ast}_{,z}-\omega_{,z}\omega^{\ast}_{,\phi})
 - A_{\phi} \cdot (\omega_{,z} \omega^{\ast} + \omega \omega^{\ast}_{,z}) +\\
 A_{z} \cdot (\omega_{,\phi} \omega^{\ast} + \omega \omega^{\ast}_{,\phi})] \bigg\} + R_{1} \cdot (A_{\phi,z} - A_{z,\phi}). \label{niezmiennik1}
 \end{gathered}
 \end{equation}

 where we assume that $R_{1}=R_{1}(\omega \omega^{\ast})$ is dependent on order parameter magnitude and this is a function, which is to be determined later.
 $R'_{1} = \frac{d R_{1}}{d (\omega \omega^{\ast})}$, so $R'_{1}$ denotes the derivative of the function $R_{1}$ with respect to its argument: $\omega\omega^{\ast}$.

 We will also use the so-called divergent invariants \cite{SOK2}-\cite{SokStSok}. Their form are: $\frac{d R_{2}}{d \phi}  = R_{2,\omega} \omega_{,\phi} + R_{2,\omega^{\ast}} \omega^{\ast}_{,\phi}$ and $\frac{d R_{3}}{d z}  = R_{3,\omega} \omega_{,z} + R_{3,\omega^{\ast}} \omega^{\ast}_{,z}$ where $R_{k}=R_{k}(\omega, \omega^{\ast}), \omega=\omega(\phi, z), \omega^{\ast}=\omega^{\ast}(\phi, z), k=1, 2$. Let us notice here that by using the lagrangian gauged on among others, divergent invariants, we can obtain the same Euler-Lagrange equations, as these ones obtained by using lagrangian ungauged on these invariants, even in that case, when the divergent invariants are not invariant under gauge transformations of the field $A_{k}, \ k=1,2$.

\section{Derivation of Bogomolny equations for the gauged G-L model in curved space}

Now, in order to apply effectively the strong necessary conditions, we need to gauge this lagrangian on the so-called invariants, and we get

\begin{equation}
\begin{gathered}
\tilde{\mathcal{L}} = \frac{a_{1} (-ih \omega_{,\phi} - \frac{eA_{\phi} \omega}{c}) (ih \omega^{\ast}_{,\phi} - \frac{eA_{\phi} \omega^{\ast}}{c})}{2m} + \frac{a_{2} (-ih \omega_{,z} - \frac{eA_{z} \omega}{c}) (ih \omega^{\ast}_{,z} - \frac{eA_{z} \omega^{\ast}}{c})}{2m}+\\
\frac{a_{3} (-ih \omega_{,\phi} - \frac{eA_{\phi} \omega}{c}) (ih \omega^{\ast}_{,z} - \frac{eA_{z} \omega^{\ast}}{c})}{2m} + \frac{a_{4} (-ih \omega_{,z} - \frac{eA_{z} \omega}{c}) (ih \omega^{\ast}_{,\phi} - \frac{eA_{\phi} \omega^{\ast}}{c})}{2m} +\\
\frac{a_{5} (-ih \omega_{,\phi} - \frac{eA_{\phi} \omega}{c})}{2m} + \frac{a_{6} (-ih \omega_{,z} - \frac{eA_{z} \omega}{c})}{2m} +
\frac{a_{7} (ih \omega^{\ast}_{,\phi} - \frac{eA_{\phi} \omega^{\ast}}{c})}{2m} + \label{lagrGLprzecech} \\
\frac{a_{8} (ih \omega^{\ast}_{,z} - \frac{eA_{z} \omega^{\ast}}{c})}{2m} + V(\omega, \omega^{\ast}) + c_{1} (A_{\phi,z} - A_{z,\phi})^{2} +\\
 \lambda R'_{1}(\omega \omega^{\ast})(i(\omega_{,\phi} \omega^{\ast}_{,z} - \omega_{,z}\omega^{\ast}_{,\phi}) -
A_{z}(\omega_{,\phi} \omega^{\ast} + \omega \omega^{\ast}_{,\phi}) + A_{\phi} (\omega_{,z} \omega^{\ast} + \omega \omega^{\ast}_{,z})) + \\
 R_{1} (A_{\phi,z} - A_{z,\phi}) + \frac{d}{d\phi} R_{2} + \frac{d}{d z}R_{3},
\end{gathered}
\end{equation}

where $c_{1}=const, R_{1} = R_{1}(\omega \omega^{\ast}), R_{2} = R_{2}(\omega, \omega^{\ast}), R_{3} = R_{3}(\omega, \omega^{\ast})$
 and the divergent invariants are: $\frac{d}{d\phi} R_{2} = R_{2,\omega}  \omega_{,\phi} + R_{2,\omega^{\ast}}  \omega^{\ast}_{,\phi}, \frac{d}{d z}R_{3} =
R_{3,\omega}  \omega_{,z} + R_{3,\omega^{\ast}}  \omega^{\ast}_{,z}$.
The functions $R_{1}, R_{2}, R_{3}$ will be determined later.

According to the strong necessary conditions, instead of considering
the Euler-Lagrange equations

\begin{gather}
\frac{d}{d r} \frac{\partial \mathcal{L}}{\partial \omega_{,r}} +
\frac{d}{d \phi} \frac{\partial \mathcal{L}}{\partial \omega_{,\phi}} + \frac{d}{d z} \frac{\partial \mathcal{L}}{\partial \omega_{,z}} -
 \frac{\partial \mathcal{L}}{\partial \omega} = 0,\\
 \frac{d}{d r} \frac{\partial \mathcal{L}}{\partial \omega^{\ast}_{,r}} +
 \frac{d}{d \phi} \frac{\partial \mathcal{L}}{\partial \omega^{\ast}_{,\phi}} + \frac{d}{d z} \frac{\partial \mathcal{L}}{\partial \omega^{\ast}_{,z}} -
 \frac{\partial \mathcal{L}}{\partial \omega^{\ast}} = 0,\\
 \frac{d}{d r} \frac{\partial \mathcal{L}}{\partial A_{r,r}} +
 \frac{d}{d \phi} \frac{\partial \mathcal{L}}{\partial A_{r,\phi}} + \frac{d}{d z} \frac{\partial \mathcal{L}}{\partial A_{r, z}} -
 \frac{\partial \mathcal{L}}{\partial A_{r}} = 0,\\
 \frac{d}{d r} \frac{\partial \mathcal{L}}{\partial A_{\phi,r}} +
 \frac{d}{d \phi} \frac{\partial \mathcal{L}}{\partial A_{\phi,\phi}} + \frac{d}{d z} \frac{\partial \mathcal{L}}{\partial A_{\phi, z}} -
 \frac{\partial \mathcal{L}}{\partial A_{\phi}} = 0,\\
  \frac{d}{d r} \frac{\partial \mathcal{L}}{\partial A_{z,r}} +
  \frac{d}{d \phi} \frac{\partial \mathcal{L}}{\partial A_{z,\phi}} + \frac{d}{d z} \frac{\partial \mathcal{L}}{\partial A_{z, z}} -
 \frac{\partial \mathcal{L}}{\partial A_{z}} = 0,
\end{gather}

we take into account just the strong necessary conditions

\begin{gather}
 \frac{\partial \tilde{\mathcal{L}}}{\partial \omega} = 0, \ \
 \frac{\partial \tilde{\mathcal{L}}}{\partial \omega^{\ast}} = 0, \label{gorne} \\
 \frac{\partial \tilde{\mathcal{L}}}{\partial \omega_{,r}} = 0, \ \
\frac{\partial \tilde{\mathcal{L}}}{\partial \omega_{,\phi}} = 0, \ \
\frac{\partial \tilde{\mathcal{L}}}{\partial \omega_{,z}} = 0, \label{dolne1} \\
\frac{\partial \tilde{\mathcal{L}}}{\partial \omega^{\ast}_{,r}} = 0, \ \
\frac{\partial \tilde{\mathcal{L}}}{\partial \omega^{\ast}_{,\phi}} = 0, \ \
\frac{\partial \tilde{\mathcal{L}}}{\partial \omega^{\ast}_{,z}} = 0, \label{dolne2} \\
\frac{\partial \tilde{\mathcal{L}}}{\partial A_{r}} = 0,
\frac{\partial \tilde{\mathcal{L}}}{\partial A_{\phi}} = 0,
\frac{\partial \tilde{\mathcal{L}}}{\partial A_{z}} = 0,\\
\frac{\partial \tilde{\mathcal{L}}}{\partial A_{r,r}} = 0,
\frac{\partial \tilde{\mathcal{L}}}{\partial A_{r,\phi}} = 0,
\frac{\partial \tilde{\mathcal{L}}}{\partial A_{r,z}} = 0,\\
\frac{\partial \tilde{\mathcal{L}}}{\partial A_{\phi,r}} = 0,
\frac{\partial \tilde{\mathcal{L}}}{\partial A_{\phi,\phi}} = 0,
\frac{\partial \tilde{\mathcal{L}}}{\partial A_{\phi,z}} = 0,\\
\frac{\partial \tilde{\mathcal{L}}}{\partial A_{z, r}} = 0,
\frac{\partial \tilde{\mathcal{L}}}{\partial A_{z,\phi}} = 0,
\frac{\partial \tilde{\mathcal{L}}}{\partial A_{z,z}} = 0,
\end{gather}

in reference to the lagrangian (\ref{lagrGLprzecech}).
We can always consider a physical system, where we assign zero value to radial component of vector potential that
is achievable in specific outside configurations of magnetic field polarizing the given system. On another hand there is always dependence of wave-function on radial position coordinate. However, we encounter
$\frac{d}{dr}=\frac{d\phi}{dr}\frac{d}{d\phi}+\frac{dz}{dr}\frac{d}{dz}$, where dependencies $\frac{d\phi}{dr}$ and $\frac{dz}{dr}$ are known due to fixed geometry of nanocable depicted in Fig.\ref{fig:enter-label}.
The equations (\ref{gorne}) - (\ref{dolne2}) have the form

\begin{equation}
\begin{gathered}
\tilde{\mathcal{L}}_{,\omega}: -\frac{a_{1} e A_{\phi} (i h \omega^{\ast}_{,\phi} - \frac{e A_{\phi} \omega^{\ast}}{c})}{2cm} -
\frac{a_{2} e A_{z} (i h \omega^{\ast}_{,z} - \frac{e A_{z} \omega^{\ast}}{c})}{2cm} -
\frac{a_{3} e A_{\phi} (i h \omega^{\ast}_{,z} - \frac{e A_{z} \omega^{\ast}}{c})}{2cm} - \\
\frac{a_{4} e A_{z} (i h \omega^{\ast}_{,\phi} - \frac{e A_{\phi} \omega^{\ast}}{c})}{2cm} +
\frac{\frac{\partial a_{5}}{\partial \omega} (-i h \omega_{,\phi} - \frac{e A_{\phi} \omega}{c})}{2cm} -
\frac{a_{5} e A_{\phi}}{2cm} +  \frac{\frac{\partial a_{6}}{\partial \omega} (-i h \omega_{,z} - \frac{e A_{z} \omega}{c})}{2cm} - \\
\frac{a_{6} e A_{z}}{2cm} + \frac{\frac{\partial a_{7}}{\partial \omega} (i h \omega^{\ast}_{,\phi} - \frac{e A_{\phi} \omega^{\ast}}{c})}{2cm} + \frac{\frac{\partial a_{8}}{\partial \omega} (i h \omega^{\ast}_{,z} - \frac{e A_{z} \omega^{\ast}}{c})}{2cm} +
\frac{\partial V}{\partial \omega} + \label{gorneI_1} \\
\lambda R''_{1} \omega^{\ast} [i (\omega_{,\phi} \omega^{\ast}_{,z} -
\omega_{,z} \omega^{\ast}_{,\phi}) - A_{\phi}(\omega_{,z} \omega^{\ast} + \omega \omega^{\ast}_{,z}) +
A_{z}(\omega_{,\phi} \omega^{\ast} + \omega \omega^{\ast}_{,\phi}) ] +\\
\lambda R'_{1} (\omega^{\ast}_{,\phi} A_{z} - \omega^{\ast}_{,z}
A_{\phi})  + R'_{1} \omega^{\ast} (A_{z,\phi} - A_{\phi,z}) + R_{2,\omega \omega} \omega_{,\phi}
R_{2,\omega \omega^{\ast}} \omega^{\ast}_{,\phi} + R_{3,\omega \omega} \omega_{,z} + R_{3,\omega \omega^{\ast}} \omega^{\ast}_{,\phi} = 0
\end{gathered}
\end{equation}

\begin{equation}
\begin{gathered}
\tilde{\mathcal{L}}_{,\omega^{\ast}}:  -\frac{a_{1} e A_{\phi} (-i h \omega_{,\phi} - \frac{e A_{\phi} \omega}{c})}{2cm} -
\frac{a_{2} e A_{z} (-i h \omega_{,z} - \frac{e A_{z} \omega}{c})}{2cm} -
\frac{a_{3} e A_{z} (-i h \omega_{,\phi} - \frac{e A_{\phi} \omega}{c})}{2cm} - \\
\frac{a_{4} e A_{\phi} (-i h \omega_{,z} - \frac{e A_{z} \omega}{c})}{2cm} +
\frac{\frac{\partial a_{5}}{\partial \omega^{\ast}} (-i h \omega_{,\phi} - \frac{e A_{\phi} \omega}{c})}{2cm} +
\frac{\frac{\partial a_{6}}{\partial \omega^{\ast}} (-i h \omega^{\ast}_{,z} - \frac{e A_{z} \omega^{\ast}}{c})}{2cm} +
\frac{\frac{\partial a_{7}}{\partial \omega^{\ast}} (i h \omega^{\ast}_{,\phi} - \frac{e A_{\phi} \omega^{\ast}}{c})}{2cm} - \\
\frac{a_{7} e A_{\phi}}{2cm} +
  \frac{\frac{\partial a_{8}}{\partial \omega^{\ast}} (i h \omega^{\ast}_{,z} - \frac{e A_{z} \omega^{\ast}}{c})}{2cm} -\label{gorneI_2}
\frac{a_{8} e A_{z}}{2cm}   +
\frac{\partial V}{\partial \omega^{\ast}} + \\
\lambda R''_{1} \omega [i (\omega_{,\phi} \omega^{\ast}_{,z} -
\omega_{,z} \omega^{\ast}_{,\phi}) - A_{\phi}(\omega_{,z} \omega^{\ast} + \omega \omega^{\ast}_{,z}) +
A_{z}(\omega_{,\phi} \omega^{\ast} + \omega \omega^{\ast}_{,\phi}) ] +\\
\lambda R'_{1} (\omega_{,\phi} A_{z} - \omega_{,z}
A_{\phi})  + R'_{1} \omega (A_{z,\phi} - A_{\phi,z}) + R_{2,\omega^{\ast} \omega} \omega_{,\phi}
R_{2,\omega^{\ast} \omega^{\ast}} \omega^{\ast}_{,\phi} + R_{3,\omega^{\ast} \omega} \omega_{,z} + R_{3,\omega^{\ast} \omega^{\ast}} \omega^{\ast}_{,\phi} = 0
\end{gathered}
\end{equation}

\begin{equation}
\begin{gathered}
\tilde{\mathcal{L}}_{,A_{\phi}}: -\frac{a_{1} e \omega (ih \omega^{\ast}_{,\phi} - \frac{e A_{\phi} \omega^{\ast}}{c})}{2cm} -\frac{a_{1} e \omega^{\ast} (-ih \omega^{\ast}_{,\phi} - \frac{e A_{\phi} \omega}{c})}{2cm} -\frac{a_{3} e \omega (ih \omega^{\ast}_{,z} - \frac{e A_{z} \omega^{\ast}}{c})}{2cm} - \label{gorneI_3} \\
-\frac{a_{4} e \omega^{\ast} (-ih \omega_{,z} - \frac{e A_{z} \omega}{c})}{2cm} - \frac{a_{5} e \omega}{2cm} - \frac{a_{7} e \omega^{\ast}}{2cm} + \lambda R'_{1} (-\omega_{,z} \omega^{\ast} - \omega^{\ast}_{,z} \omega) = 0,
\end{gathered}
\end{equation}

\begin{equation}
\begin{gathered}
\tilde{\mathcal{L}}_{,A_{z}}: -\frac{a_{2} e \omega (ih \omega^{\ast}_{,z} - \frac{e A_{z} \omega^{\ast}}{c})}{2cm} -\frac{a_{2} e \omega^{\ast} (-ih \omega^{\ast}_{,z} - \frac{e A_{z} \omega^{\ast}}{c})}{2cm} -\frac{a_{3} e \omega^{\ast} (ih \omega^{\ast}_{,\phi} - \frac{e A_{\phi} \omega}{c})}{2cm} - \label{gorneI_4} \\
-\frac{a_{4} e \omega (ih \omega^{\ast}_{,\phi} - \frac{e A_{\phi} \omega^{\ast}}{c})}{2cm} - \frac{a_{6} e \omega}{2cm} - \frac{a_{8} e \omega^{\ast}}{2cm} + \lambda R'_{1} (\omega_{,\phi} \omega^{\ast} + \omega^{\ast}_{,\phi} \omega) = 0
\end{gathered}
\end{equation}

\begin{equation}
\begin{gathered}
\tilde{\mathcal{L}}_{,\omega_{,\phi}}: -\frac{ia_{1} h (ih \omega^{\ast}_{,\phi} - \frac{e A_{\phi} \omega^{\ast}}{c})}{2m} -
\frac{ia_{3} h (ih \omega^{\ast}_{,z} - \frac{e A_{z} \omega^{\ast}}{c})}{2m} - \label{dolneI_1} \\
 \frac{ia_{5} h}{2m} + \lambda R'_{1} (i\omega^{\ast}_{,z} + A_{z} \omega^{\ast}) + R_{2,\omega} = 0
\end{gathered}
\end{equation}

\begin{equation}
\begin{gathered}
\tilde{\mathcal{L}}_{,\omega_{,z}}: -\frac{ia_{2} h (ih \omega^{\ast}_{,z} - \frac{e A_{z} \omega^{\ast}}{c})}{2m} -
\frac{ia_{4} h (ih \omega^{\ast}_{,\phi} - \frac{e A_{\phi} \omega^{\ast}}{c})}{2m} - \label{dolneI_2} \\
 \frac{ia_{6} h}{2m} + \lambda R'_{1} (-i\omega^{\ast}_{,\phi} - A_{\phi} \omega^{\ast}) + R_{3,\omega} = 0
\end{gathered}
\end{equation}
\begin{equation}
\begin{gathered}
\tilde{\mathcal{L}}_{,\omega^{\ast}_{,\phi}}: \frac{ia_{1} h (-ih \omega_{,\phi} - \frac{e A_{\phi} \omega}{c})}{2m} +
\frac{ia_{4} h (-ih \omega_{,z} - \frac{e A_{z} \omega}{c})}{2m} + \label{dolneI_3} \\
 \frac{ia_{7} h}{2m} + \lambda R'_{1} (-i\omega_{,z} + A_{z} \omega) + R_{2,\omega^{\ast}} = 0
\end{gathered}
\end{equation}

\begin{equation}
\begin{gathered}
\tilde{\mathcal{L}}_{,\omega^{\ast}_{,z}}: \frac{ia_{2} h (-ih \omega_{,z} - \frac{e A_{z} \omega}{c})}{2m} +
\frac{ia_{3} h (-ih \omega_{,\phi} - \frac{e A_{\phi} \omega}{c})}{2m} + \label{dolneI_4} \\
 \frac{ia_{8} h}{2m} + \lambda R'_{1} (i\omega_{,\phi} - A_{\phi} \omega) + R_{3,\omega} = 0
\end{gathered}
\end{equation}

\begin{gather}
\tilde{\mathcal{L}}_{,A_{\phi,z}}: -2c_{1} (A_{z,\phi} - A_{\phi, z}) - R_{1} = 0,  \label{dolneI_A1}\\
\tilde{\mathcal{L}}_{,A_{z,\phi}}: 2c_{1} (A_{z,\phi} - A_{\phi, z}) + R_{1} = 0. \label{dolneI_A2}
\end{gather}

Now, we have to make the equations (\ref{gorneI_1}) - (\ref{dolneI_4}) self-consistent. Thus, the reduction of
the number of independent equations by an appropriate choice of the functions $R_{k}, (k =1, 2, 3)$, is necessary.
Usually, such ansatzes exist only for some special $V(\omega, \omega^{\ast})$. Hence, in most cases of $V(\omega,\omega^{\ast})$ for many nonlinear field models, the reduction of the system of corresponding dual equations, to Bogomolny equations, is impossible.

At the beginning, we make the equations
(\ref{dolneI_1}) - (\ref{dolneI_4}) as linear dependent. In this order we must put
$a_{m} = a_{4} \ (m=1,2,3)$ and $a_{5}(\omega, \omega^{\ast})=-a_{8}(\omega, \omega^{\ast}),
a_{6}(\omega, \omega^{\ast})=-a_{8}(\omega, \omega^{\ast}), a_{7}(\omega, \omega^{\ast})=a_{8}(\omega, \omega^{\ast})$
and $R_{1} \equiv c_{R_{1}}=const, R_{2,\omega}=const, R_{2,\omega^{\ast}}=const, R_{3,\omega}=const, R_{3,\omega^{\ast}}=const$.
Then, we have the candidates for the wanted Bogomolny equations for
this model

\begin{gather}
\omega_{,\phi} + \omega_{,z} = \frac{i}{a_{4} ch} (A_{\phi} a_{4} e \omega + A_{z} a_{4} e \omega - a_{8}(\omega, \omega^{\ast}) c),
\label{BPS1}\\
\omega^{\ast}_{,\phi} + \omega^{\ast}_{,z} = -\frac{i}{a_{4} ch} (A_{\phi} a_{4} e \omega^{\ast} + A_{z} a_{4} e \omega^{\ast} - a_{8}(\omega, \omega^{\ast}) c), \label{BPS2}\\
2c_{1} (A_{z,\phi} - A_{\phi, z}) + c_{R_{1}} = 0, \label{BPS3}
\end{gather}

We eliminate all terms including the derivatives of $\omega, \omega^{\ast}, A_{\phi}, A_{z}$ in (\ref{gorneI_1}) - (\ref{gorneI_4})
and we get the system of partial differential equations for $V$:

\begin{gather}
\frac{1}{a_{4}m}(V_{,\omega} a_{4} m +  a_{8, \omega}) = 0, \\
\frac{1}{a_{4}m}(V_{,\omega^{\ast}} a_{4} m +  a_{8, \omega^{\ast}}) = 0.
\end{gather}

Thus, we obtain the condition for $V$:

\begin{equation}
V(\omega, \omega^{\ast}) = -\frac{a^{2}_{8}(\omega, \omega^{\ast})}{2a_{4}m} + c_{2},
\end{equation}

where $c_{2}=const$.

When this condition is satisfied, the equations (\ref{BPS1}) - (\ref{BPS3}) are the wanted Bogomolny equations (BPS equations) for this model.

\section{Conclusions}

 We have derived the Bogomolny equations (BPS equations), for the gauged Ginzburg-Landau model in curved space,  given by the lagrangian (\ref{lagrGL}).
 The density of the topological charge for the solutions
 of the derived Bogomolny equations, is given by

 \begin{gather}
 \lambda R'_{1}(\omega \omega^{\ast})(i(\omega_{,\phi} \omega^{\ast}_{,z} - \omega_{,z}\omega^{\ast}_{,\phi}) -
A_{z}(\omega_{,\phi} \omega^{\ast} + \omega \omega^{\ast}_{,\phi}) + A_{\phi} (\omega_{,z} \omega^{\ast} + \omega \omega^{\ast}_{,z})) + \\
 R_{1} (A_{\phi,z} - A_{z,\phi}),
 \end{gather}

 where $R_{1}=c_{R_{1}}=const$, so this density is reduced to

 \begin{gather}
 c_{R_{1}} (A_{\phi,z} - A_{z,\phi}). \label{gestLad_poleMagn}
 \end{gather}

Hence, we have the direct connection between the density of topological charge and the magnetic field (given just
by (\ref{gestLad_poleMagn})). In such a case the magnetic field is space-independent, what experimentally corresponds to spiral-like superconducting cable placed in side solenoid generating uniform magnetic field.

\section{Acknowledgment}

The computations were carried out, by using Waterloo MAPLE Software, owing to the financial support, provided by The Pedagogical University of Cracow, within the University Research Projects (the numbers: WPBU/2022/04/00319 and WPBU/2023/04/00009).


\begin{thebibliography}{100} 
\bibitem{Bogomolny}
E.B.Bogomolny, \emph{Stability of Classical Solutions}, Sov. J. Nucl. Phys., 24:449, 1976

\bibitem{BystrovEtAl}
A. S. Bystrov, A. S. Mel’nikov, D. A. Ryzhov, I. M. Nefedov, I. A. Shereshevskii, and P. P.
Vysheslavtsev, \emph{Singular and nonsingular vortex structures in high-temperature superconductors}, Mod. Phys. Lett. B, 17:621, 2003

\bibitem{ContiOttoSerfaty}
S.Conti, F.Otto, S.Serfaty, \emph{Branched Microstructures in the Ginzburg-Landau Model of Type-I Superconductors}, SIAM J. Math. Anal., 48:2994, 2015, arXiv:1507.00836

\bibitem{Morandi} G.Morandi, \emph{The Role of Topology in Classical and Quantum Physics}, Springer-Verlag, 1992

\bibitem{BPST}
A.A.Belavin, A.M.Polyakov, A.S.Schwartz, Y.S.Tyupkin, \emph{Pseudoparticle solutions of the Yang-Mills equations}, Phys. Lett., B59:85, 1975
\bibitem{Hosoya}
A.Hosoya, \emph{On Vanishing of Energy-Momentum Tensor for a Class of Instanton-Like Solutions}, Prog. Theor. Phys., 59:1781, 1978
\bibitem{Adam_Santamaria}
C. Adam and F. Santamaria, \emph{The first-order Euler-Lagrange equations and some of
their use}, JHEP, 2016:047, 2016. arXiv:1609.02154.
\bibitem{MantonSutcliffe}
N. Manton and P. Sutcliffe, \emph{Topological Solitons}, Cambridge University Press, 2004
\bibitem{Meissner2013}
K.A.Meissner, \emph{The Classical Field Theory}, PWN, 2013 in Polish
\bibitem{Bialynicki}
I. Bialynicki-Birula, \emph{On the stability of solitons}, Nonlinear Problems in
Theoretical Physcis, 1978, Vol 98 of Lect. Notes Phys., page 15, Springer Basel, 1979
\bibitem{Albert} J.Albert,\emph{The Abrikosov Vortex in Curved Space}, JHEP, 2021.
\bibitem{AkkermansMallick} E. Akkermans, K. Mallick,\emph{Geometrical Description of Vortices in Ginzburg-Landau Billiards }, Vol.199, 1999
\bibitem{BystrovEtal} A.S.Bystrov, A.S.Mel'nikov, D.A.Ryzhov, I.M.Nefedov, I.A.Shereshevskii, P.P.Vysheslavtsev,\emph{Singular and Nonsingular Vortex Structures in High-Temperature Superconductors}, Modern Physics Letters B, Vol.17, 2003
\bibitem{Jaykka} J.Jaykka, J.Palmu,\emph{Knot solitons in modified Ginzburg-Landau model},Vol.83, Physical Review D, 2011
\bibitem{SandierSerfaty} E. Sandier and S. Serfaty, \emph{Gamma-convergence of gradient flows with applications to Ginzburg–Landau}, Comm. Pure Appl. Math., Vol.4, Analysis and PDE, 2011
\bibitem{Contatto} F. Contatto, \emph{Integrable Abelian vortex-like solitons} , Physical Letters B, Vol.768, 2017


\bibitem{SOK1979} K.Sokalski, \emph{Instantons in Anisotropic Ferromagnets}, Acta Phys. Pol., Vol.A56, 1979


\bibitem{Felsager}
B.Felsager,\emph{Geometry, Particles  and Fields}, Springer, 1998


\bibitem{SOK22}
K. Sokalski, T. Wietecha, and Z. Lisowski, \emph{A Concept of Strong Necessary Condition in Nonlinear Field Theory}, Acta Phys. Pol., B32:2771, 2001.
\bibitem{SOK2a}
K. Sokalski, T. Wietecha, and Z. Lisowski, \emph{Unified variational approach to the Bäcklund transformation and the Bogomolny decomposition}, Int. J. Theor. Phys., Group Theory and Nonl. Optics,
NOVA, 9:331, 2002

\bibitem{SOK1993}
P.T.Jochym, K.Sokalski,
Journal of Physics A, \emph{Mathematical and General Variational approach to the Bogomolny separation}, J. Phys. A: Math. Gen., 26:3837, 1993.

\bibitem{SOK2}
K.Sokalski, T.Wietecha, Z.Lisowski, \emph{A Concept of Strong Necessary Condition in Nonlinear Field Theory}, Acta Phys. Pol., B32:17, 2001

\bibitem{ST2015}
L.T.Stepien, \emph{Bogomolny equation for the BPS Skyrme model from strong necessary conditions}, J. Phys. A, 49:175202, 2016
\bibitem{ST11}
L.T.Stepien, J. Phys. A, 51:015208, 2018

\bibitem{Weinan}
E. Weinan, Physica D: Nonl. Phen. Elsevier, 77:383, Dynamics of vortices in Ginzburg-Landau theories with applications to superconductivity, 1994
\bibitem{Yang}
Y. Yang. Z. Angew, \emph{On the existence of multivortices in a generalized Bogomol'nyi system}, Math. Phys., 43:677, 1992

\bibitem{PeninWeller},
A. Penin, Q. Weller, \emph{A Theory of Giant Vortices}, JHEP, 2021:056, 2021, arXiv:2105.1213

\bibitem{Serfaty}
S.Serfaty, \emph{Emergence of the Abrikosov lattice in several models with two dimensional Coulomb
interaction}, In R. Latala, A. Rucinski, P.Strzelecki, J.Swiatkowski, D. Wrzosek, and P. Zakrzewski,
editors, European Congress of Mathematics, Cracow, 2-7 July 2012, page 119, 2014.

\bibitem{SOK1981}
K. Sokalski, \emph{Dynamical stability of instantons}, pages: 102-103 , Phys. Lett., Vol. A 81,1981
\bibitem{SOK1a}
K.Sokalski, Acta Phys. Pol., Vol.A 65, 1984
\bibitem{SokStSok}
K.Sokalski, L.Stepien, D. Sokalska, J. Phys, Vol.A35, 2002
\bibitem{SOK2005}
K.Sokalski, T.Wietecha, D.Sokalska, J. Nonl. Math. Phys., 12:31, 2005
\bibitem{SOK2009}
L. Stepien, D.Sokalska, K.Sokalski, \emph{The Bogomolny decomposition for systems of two generalized nonlinear partial differential equations of the second order}, J. Nonl. Math. Phys., Vol.16, 2009

\bibitem{ST12}
L.T.Stepien. J. Phys. A, Vol.51, \emph{Bogomolny equations in certain generalized baby BPS Skyrme models}, 015208, 2018.

\bibitem{ST3} L.T.Stepien, \emph{Bogomolny equations for the BPS Skyrme models with impurity}, Journal of High Energy Physics, Vol. 140, 2020.

\bibitem{ST1}
L.Stepien, \emph{Bogomolny decomposition in the context of the concept of strong necessary conditions},
PhD thesis defended at Jagiellonian University, 2003

\bibitem{Pomorski} K.Pomorski, \emph{Analytical view on topological defects of superconducting order parameter in various topologies of nanowires with focus on quantum detectors and Josephson junctions} , Molecular Crystals and Liquid Crystals, Vol.750, 2022
\bibitem{Curved} K.Pomorski, Arxiv:2112.13470, \emph{Electrostatically interacting Wannier qubits in curved space}, 2021
\end{thebibliography}
\end{document}